\documentclass[twocolumn,preprintnumbers,prb,aps,amssymb,superscriptaddress]{revtex4}

\usepackage{graphicx}
\usepackage{dcolumn}
\usepackage{bm}

\begin{document}

\title{Local tunneling probe of (110) Y$_{0.95}$Ca$_{0.05}$Ba$_2$Cu$_3$O$_{7-\delta}$ thin films in a magnetic field}

\author{J. H. Ngai}
 \affiliation{Department of Physics, University of Toronto, 60 St. George Street, Toronto, ON  M5S1A7, Canada}
\author{R. Beck}
 \affiliation{Department of Physics and Astronomy, Raymond and Beverly Sackler Faculty of Exact Sciences, Tel Aviv University, 69978 Tel Aviv, Israel}
\author{G. Leibovitch}
 \affiliation{Department of Physics and Astronomy, Raymond and Beverly Sackler Faculty of Exact Sciences, Tel Aviv University, 69978 Tel Aviv, Israel}
\author{G. Deutscher}
 \affiliation{Department of Physics and Astronomy, Raymond and Beverly Sackler Faculty of Exact Sciences, Tel Aviv University, 69978 Tel Aviv, Israel}
\author{J. Y.T. Wei}%
 \affiliation{Department of Physics, University of Toronto, 60 St. George Street, Toronto, ON  M5S1A7, Canada}

\date{\today}

\begin{abstract}

Scanning tunneling spectroscopy was performed on (110)-oriented thin films of Ca-overdoped Y$_{0.95}$Ca$_{0.05}$Ba$_2$Cu$_3$O$_{7-\delta}$ at 4.2K, to probe the local evolution of Andreev$-$Saint-James surface states in a $c$-axis magnetic field. In zero field, we observed conductance spectra with spontaneously-split peaks and spectra with unsplit zero-bias peaks. The former showed enhanced splitting with field, and the latter showed threshold splitting above finite fields. Although both field evolutions can be described in terms of screening and orbital supercurrents, within the framework of $d\pm i\alpha$ pairing ($d$=$d_{x^2-y^2}$; $\alpha$=$d_{xy}$,$s$), the enhanced splitting is consistent with only the $d$+$i\alpha$ state. Our results have direct implications on the stability of broken time-reversal symmetry in cuprate superconductors.

\end{abstract}

\maketitle

The predominance of $d$-wave pairing symmetry in the high critical-temperature ($T_c$) superconducting cuprates is responsible for a wealth of unconventional phenomena \cite{Tsuei_Kirtley_revu,vanHarlingen_revu}. Of particular interest is the formation of Andreev$-$Saint-James (ASJ) surface states which are manifested as zero-bias peaks in the conductance spectra taken primarily on (110) tunnel junctions \cite{Deutscher_revu}. These low-energy ASJ states arise from the constructive interference between time-reversed quasiparticles about the $d$-wave line nodes, thus providing direct information about the high-$T_c$ order parameter (OP) \cite{CRHu,TK95,WeiPRL98,JNgai05}. In the case of YBa$_2$Cu$_3$O$_{7-\delta}$ (YBCO), much attention has been focused on the peak splitting observed in zero field at low temperatures \cite{Deutscher_revu}, primarily on overdoped samples \cite{Dagan01}. This spontaneous peak splitting has generally been interpreted as evidence for a complex $d\pm i\alpha$ OP ($d$=$d_{x^2-y^2}$; $\alpha$=$d_{xy}$,$s$) with intrinsically broken time-reversal symmetry (BTRS) \cite{Covington_Greene}.  Alternatively, there have also been theoretical suggestions that the spontaneous peak splitting can arise extrinsically, from either electron-hole asymmetry, multiband effects or impurity perturbation \cite{Golubov_Tafuri,Tanuma_multiband,Asano}.

The occurrence of intrinsic BTRS in the thermodynamic ground state of a cuprate superconductor would have profound theoretical implications on the high-$T_c$ pairing mechanism \cite{Vojta_Sachdev}.  For YBCO it is believed that a spontaneous orbital supercurrent, associated with an $i\alpha$ OP component, should exist on (110) surfaces.  However, this spontaneous current has not been directly observed.  In earlier tunneling studies of YBCO, the existence of this spontaneous current was inferred through a Doppler interpretation of the magnetic-field evolution of the ASJ peak splitting \cite{Aprili_Greene,Fogelstrom_PRL}.  This interpretation assumes that the field-induced screening current enhances the peak splitting through Doppler-like energy shifts of the ASJ surface states.  More recent tunneling experiments on YBCO have shown that the rate of peak splitting with decreasing field is invariant as screening is suppressed with film thinness, suggesting that a non-Doppler mechanism for the field splitting is also at play \cite{Beck04,Elhalel07,Laughlin98,Balatsky00}.

Another puzzling question about the Doppler interpretion is why, of all the tunneling data reported on YBCO to date, only field enhancement of the peak splitting has been seen.  Since the two possible states of $d\pm i\alpha$ correspond to oppositely orbiting supercurrents, their spectral manifestations in a $c$-axis field should be distinctly different.  Namely, the peak splitting is expected to increase/decrease if the screening and orbital supercurrents are parallel/antiparallel.  Detailed calculations of the peak splitting have indicated a variety of field evolutions, i.e. asymmetries, thresholds and discontinuities, depending on the relative strength of $\alpha$ to $d$ \cite{Fogelstrom04}. Of these possibilities, only the threshold effect has been reported, and only in underdoped YBCO \cite{Leibovitch08}. It should be noted that all of these prior in-field measurements were done on \emph{macroscopic} planar junctions, which are not sensitive to local variations in the orbital supercurrents. Recent Josephson tunneling and magneto-optics experiments on Sr$_2$RuO$_4$ have indicated that a complex OP may actually be granularized into time-reversed domains, particularly in the case of chiral symmetry \cite{vanHarlingen_SrRuO,Kapitulnik_SrRuO,Tanaka_Sigrist}. 

This Letter reports on the first \emph{microscopic} examination of these issues, using scanning tunneling spectroscopy (STS) on (110)-oriented thin films of Ca-overdoped Y$_{0.95}$Ca$_{0.05}$Ba$_2$Cu$_3$O$_{7-\delta}$ (Ca-YBCO). By virtue of its nanometer junction size, STS provides a direct probe of the high-$T_c$ OP at coherence-length scales. In zero field at 4.2K, we observed conductance spectra with spontaneously-split peaks as well as spectra with unsplit zero-bias peaks. The former spectra showed enhanced splitting in an increasing $c$-axis field, and the latter spectra showed threshold splitting above finite fields.  We analyze these spectral evolutions versus field as the combined effects of screening and orbital supercurrents. Although both field evolutions can be described within the framework of $d\pm i\alpha$ pairing symmetry, the former is consistent with only the $d$+$i\alpha$ state. The apparent absence of the $d$-$i\alpha$ state has direct implications on the stability of BTRS in cuprate superconductors.

\begin{figure}[t]
\centering
\includegraphics[width=8.5cm] {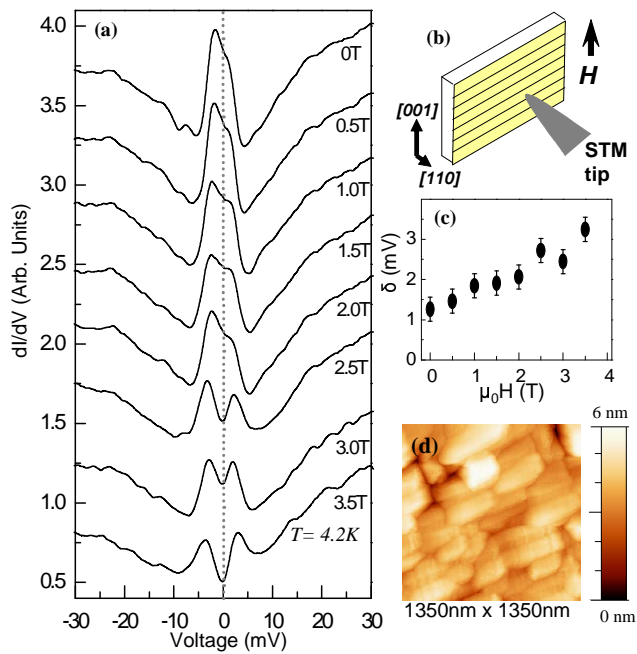}
\caption{\label{71106_Fig1} (color online) (a) Magnetic field evolution of a spontaneously-split $dI/dV$ peak spectrum taken by STS with a Pt-Ir tip on a (110)-oriented Y$_{0.95}$Ca$_{0.05}$Ba$_2$Cu$_3$O$_{7-\delta}$ thin film at 4.2K. The spectra, staggered for clarity, show the peak splitting $\delta$ increasing with the $c$-axis field $H$. (b) Schematic of the in-field STM orientation. (c) Plot of $\delta$ versus $H$. (d) STM image of a film showing strongly-oriented surface morphology.}
\end{figure}

The overdoped Ca-YBCO films measured in this work were grown by DC sputtering on (110)-oriented SrTiO$_{3}$ (STO) substrates, with a buffer layer of PrBa$_2$Cu$_3$O$_{7-\delta}$ to ensure proper epitaxy. The films were typically 160nm thick and showed  $T_c\approx$ 85K, consistent with Ca overdoping \cite{Bottger}. The films were transported in airtight containers and reannealed in flowing O$_2$ before being cooled with ultra-pure He gas to 4.2K. The STS apparatus was specially designed to sit horizontally in a $c$-axis magnetic field, as shown in Figure \ref{71106_Fig1}(b).  STS measurements were made with Pt-Ir tips by suspending the feedback and measuring the tunneling current $I$ versus bias voltage $V$. STM setpoints of 1nA and 100mV were typically used.  For each STS junction, fifty $I$-$V$ sweeps were averaged and then differentiated to give the conductance $dI/dV$ spectrum at each field.  Ramping between field points was done very slowly at $\sim$0.19T/min.  Since our STM was made entirely from non-magnetic materials, the stability of our junctions was mainly limited by piezoelectric drift, estimated to be less than 1nm at 4.2K from STM images taken on NbSe$_2$.

In zero magnetic field, we observed $dI/dV$ spectra with spontaneously-split peaks as well as spectra with unsplit peaks, varying over distances as short as $\sim$3nm. Figure \ref{71106_Fig1}(a) shows the field evolution of a spontaneously-split peak spectrum.  The magnitude of the peak splitting $\delta(H)$, defined relative to zero bias, is seen to increase from 1.3mV to 3.3mV as the field $H$ is increased from 0T to 3.5T.  This overall trend of $\partial\delta/\partial H>0$ is summarized in Fig.\ref{71106_Fig1}(c).  The relative heights of the split peaks appear to change with the field; the negative-bias peak is taller than the positive-bias peak at low fields, but this asymmetry reverses polarity as $H$ reaches $\approx$ 3T.  Although such peak-height asymmetry was generally seen on our films, its bias polarity and field dependence varied spatially.  Also worth noting in the spectra of Fig.\ref{71106_Fig1}(a) are the gap-like features denoted by a slope change in $dI/dV$ near $\pm$18mV, which become less distinct from the spectral background with increasing field. These gap-like structures were also seen in planar-junction studies \cite{Sharoni}, and could be attributed to $ab$-plane tunneling contributions associated with surface roughness.  Figure \ref{71106_Fig1}(d) shows the STM topograph of a typical film surface, with similar morphology as seen in previous studies \cite{Sharoni}. 

\begin{figure}[t]
\centering
\includegraphics[width=8.5cm] {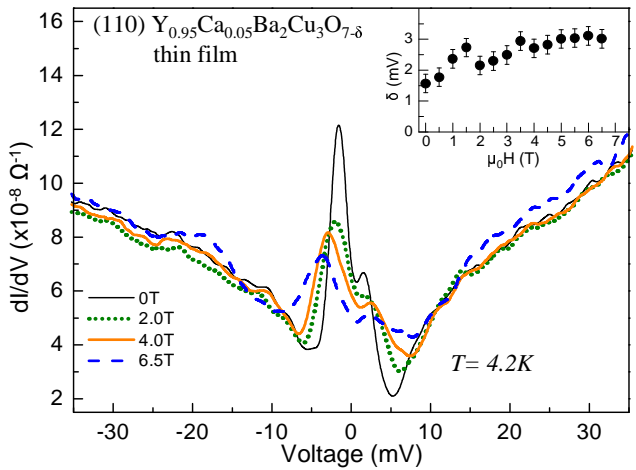}
\caption{\label{62906_Fig2} (color online) Magnetic field evolution of a spontaneously-split $dI/dV$ peak spectrum taken by STS with a Pt-Ir tip on a (110)-oriented  Y$_{0.95}$Ca$_{0.05}$Ba$_2$Cu$_3$O$_{7-\delta}$ thin film at 4.2K. The peak splitting $\delta$ observed at 0T (black) increases with the $c$-axis field $H$: 2T (green), 4T (red) and 6.5T (blue).  The inset plots $\delta(H)$ showing the overall trend.}
\end{figure}

Figure \ref{62906_Fig2} shows the field evolution of another spontaneously-split peak spectrum, with a larger $\delta$(0) and greater height asymmetry. Here the peak splitting is also enhanced by the field, as shown by $\delta (H)$ plotted in the inset, but the polarity of the asymmetry does not change with field. It should be noted that $\delta$(0) as large as 2mV were seen across our film surfaces. Nevertheless, in virtually all junctions showing spontaneous peak splitting, an overall trend of $\partial\delta/\partial$$H$$>$0 was seen, regardless of the magnitude of $\delta$(0). For several junctions we also noticed small discontinuities in $\delta (H)$ at varying levels of $H$. For the spectra shown in Fig.\ref{62906_Fig2}, a discontinuity occurs between 1.5T and 2.0T, as is visible in the inset.

\begin{figure}[t]
\centering
\includegraphics[width=8.5cm] {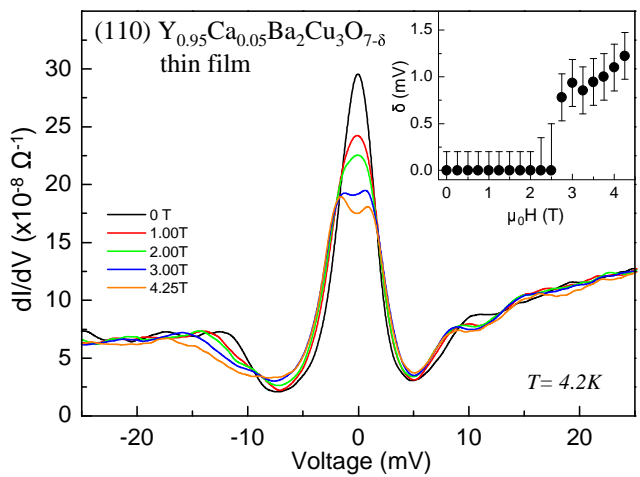}
\caption{\label{112806_Fig3} (color online) Magnetic field evolution of an originally-unsplit $dI/dV$ peak spectrum taken by STS with a Pt-Ir tip on a (110)-oriented Y$_{0.95}$Ca$_{0.05}$Ba$_2$Cu$_3$O$_{7-\delta}$ film at 4.2K.  As the $c$-axis field $H$ is applied, the zero-bias peak dips and broadens before splitting to $\approx$0.8mV at a threshold $\approx$2.5T, as shown by the peak-spitting $\delta(H)$ plot in the inset.}
\end{figure}

Figure \ref{112806_Fig3} shows the field evolution of a peak spectrum with no spontaneous splitting.  As field is increased, the peak drops in height and broadens in width before splitting to $\approx$ 0.8mV at a threshold of $\approx$ 2.5T.  This threshold behavior is clearly visible in the $\delta(H)$ plotted in the inset. The field thresholds we observed showed variation across the film surfaces. It is worth noting that the gap-like features in these spectra occur near $\pm$10mV in zero field and appear to shift asymmetrically in voltage with increasing field. Furthermore, the unsplit peaks observed in these spectra tend to be taller than the spontaneously-split peaks shown in Fig.\ref{71106_Fig1}(a) and Fig.\ref{62906_Fig2}. In general, all the spectral peaks we observed are taller than those reported in planar junctions measurements, consistent with the microscopic nature of our STS probe.

Our STS data are analyzed in the framework of $d\pm i\alpha$ pairing symmetry with intrinsic BTRS, to elucidate the origin of the spontaneous peak splitting. In the presence of an $i\alpha$ OP component on the (110) surface, the ASJ surface states are expected to shift by $\pm\alpha$ about zero energy, giving rise to a spontaneous peak splitting $\delta$(0)=$\alpha$ \cite{Fogelstrom_PRL,Zhu99,Tanuma99}. At zero field, the variation in $\delta$(0) seen on our films indicates that $\alpha$ varies spatially. Previous tunneling studies have found that $\delta$(0) tends to increase with overdoping \cite{Dagan01,Sharoni}, thus the spatial variation we observed could be due to doping inhomogeneity.  Alternatively, this variation could be due to electronic inhomogeneity, which may arise from proximity to a quantum critical point \cite{Dagan01,Vojta_Sachdev}.  Here it is worth noting that the combination of zero-bias peak, smaller gap size and threshold splitting seen in Fig.\ref{112806_Fig3} resembles the planar-junction data taken on underdoped YBCO \cite{Leibovitch08}.  

In the present literature on tunneling in YBCO, the existence of a spontaneous orbital current associated with an $i\alpha$ OP component is inferred from the $\delta(H)$ evolution \cite{Aprili_Greene,Fogelstrom_PRL}. In this Doppler scenario, the screening current combines with the orbital supercurrent to shift the energies of the ASJ surface states. At low fields, the screening current alone is expected to cause a linear shift $\vec{v_F}\cdot\vec{p_s} $cos$\phi_c$ \cite{Fogelstrom_PRL}, where $\vec{v_F}$ is the Fermi velocity, $\vec{p_s}$ is the superfluid momentum and $\phi_c$ models the tunneling cone \cite{Tunnelcone}.  A more explicit calculation by Fogelstr\"om \emph{et al.} has predicted a variety of nonlinear $\delta (H)$ behaviors, i.e. field asymmetries, thresholds and discontinuities, depending on the relative strength of $\alpha$ to $d$ \cite{Fogelstrom04}. These nonlinear behaviors arise as the screening and orbital currents couple to modify both the amplitude and phase of $i\alpha$ on the (110) surface. The discontinuities and threshold effects we observed (Figs. 2 and 3) are consistent with this scenario. It is worth noting that these nonlinear effects varied spatially in our measurements, giving evidence for inhomogeneities in the OP symmetry. These observations attest to the microscopic nature of our STS junctions, which can resolve spatial variations that are presumably averaged out in the macroscopic planar junctions. 

\begin{figure}[t]
\centering
\includegraphics[width=8.5cm] {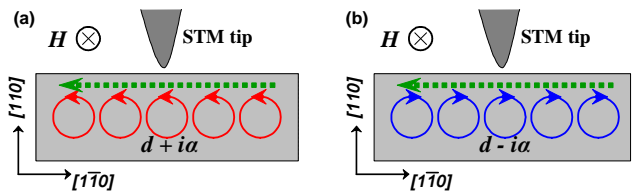}
\caption{\label{Domains_Fig4} (color online) Heuristic illustration of STS on the (110) surface of a $d\pm i\alpha$ superconductor in a $c$-axis magnetic field. Field-induced screening currents are represented by the dashed arrow, and spontaneous orbital currents by the loops. (a) For $d$+$i\alpha$, screening and orbital currents flow parallel near the surface, causing the spontaneously-split peaks to open up further with field. (b) For $d$-$i\alpha$, screening and orbital currents flow antiparallel, causing the peaks to close in with field.}
\end{figure}

In principle, the $d$+$i\alpha$ and $d$-$i\alpha$ states can be differentiated by considering the relative orientation between the screening and orbital supercurrents. As illustrated in Figure \ref{Domains_Fig4}, $d$+$i\alpha$ ($d$-$i\alpha$) corresponds to parallel (antiparallel) flow between the field-induced and spontaneous supercurrents near the (110) surface.  The model calculations of Ref.\cite{Fogelstrom04} have shown that the ASJ surface states for $d$+$i\alpha$ and $d$-$i\alpha$ tend to be Doppler-shifted in opposite directions about zero energy, causing the spontaneously-split peaks to either open up further or close in instead \cite{Fogelstrom04}. Remarkably, all the spontaneously-split peaks in our measurements showed enhanced splitting with field ($\partial\delta/\partial$$H$$>$0), consistent with only the $d$+$i\alpha$ state. This observation is puzzling, since $d$+$i\alpha$ and $d$-$i\alpha$ are degenerate in zero field, and thus half of spontaneously-split peak spectra should show $\partial\delta/\partial$$H$$<$0. In this regard, the calculations of Ref.\cite{Fogelstrom04} also indicated that parallel flow tends to be energetically favorable over antiparallel flow, thereby preferring $d$-$i\alpha$ to ``flip'' into $d$+$i\alpha$ above some threshold field $H^*$. Thus the field increments used in our measurements provide an empirical upper estimate on $H^*$ $\approx$ 0.5T, which is within the theoretical range of 0.03T$-$0.63T estimated for the $d\pm i\alpha$ scenario \cite{Fogelstrom04}.  In a followup study we will more precisely measure $H^*$ by applying finer field increments in the lower field regime. 
 
The relative stabilities of the $d$+$i\alpha$ and $d$-$i\alpha$ states in a magnetic field can also be considered in the context of OP domains \cite{vanHarlingen_SrRuO,Kapitulnik_SrRuO}.  In this scenario, the OP is granularized into $d$+$i\alpha$ and $d$-$i\alpha$ domains, whereby the spontaneous orbital supercurrents from opposing domains cancel when spatially averaged. As a result, it would be difficult to detect such spatially-varying orbital supercurrents through macroscopic techniques. Because of its small junction size, our STS measurement provides a microscopic probe of the OP, presumably within single domains. Therefore, the $H^*$ estimated above can alternatively be viewed as an upper limit on the field required to induce reversal between $d$-$i\alpha$ and $d$+$i\alpha$ domains.     

The spontaneous peak splitting we observed indicates that $\alpha$ is $\approx$ 10\% of $d$ on our overdoped Ca-YBCO (110) films at zero field. This observation seems exceptional since bulk measurements have suggested $\alpha$ to be less than $\approx$ 5\% of $d$ throughout the superconducting regime in the cuprate phase diagram over temperature and doping \cite{Tsuei_universal,Neils_vanHarlingen,Sutherland}. However, the possibility of OP domains was not explicitly considered in the analysis of these bulk experiments, thus a re-examination of this possibility would be important. Here we emphasize that it is entirely conceivable for $d$-wave pairing to be robust in the bulk and still have an $i\alpha$ OP component near (110) surfaces due to their pair-breaking nature \cite{Fogelstrom_PRL,Zhu99,Tanuma99}. 

Our analysis indicates that, in order to attribute the spontaneous peak splitting we observed to an intrinsic $i\alpha$ OP component, it would be necessary to invoke field-induced flipping of $d$-$i\alpha$ to $d$+$i\alpha$.  Alternative theoretical calculations have suggested that spontaneous peak splitting could also arise extrinsically, from either electron-hole asymmetry, multiband coupling or impurity perturbation \cite{Golubov_Tafuri,Tanuma_multiband,Asano}.  First, electron-hole asymmetry, which is inherent in the Fermi-surface topology of cuprates \cite{Wei_prb}, can cause retroflected quasielectrons and quasiholes to experience different barrier strengths, thereby upsetting their constructive interference \cite{Golubov_Tafuri}.  Second, there is experimental evidence to suggest that the plane and chain bands of YBCO are coupled \cite{JHNgai07}, thus providing a multiband mechanism for the spontaneous peak splitting \cite{Tanuma_multiband}. Although both of these mechanisms are plausible, at present it is unclear whether they can also account for the temperature and field dependences of the peak splitting \cite{Dagan01,Covington_Greene,Aprili_Greene}. Lastly, numerical calculations have shown that impurities can also produce spontaneous and field-enhanced peak splitting \cite{Asano}. However, the $\delta (H)$ thresholds and discontinuities seen in our measurements were not predicted by these calculations, thus the impurity scenario seems unlikely to explain our data.  Further theoretical work is needed to elucidate these possibilities.  

In summary, we have performed scanning tunneling spectroscopy on (110) Y$_{0.95}$Ca$_{0.05}$Ba$_2$Cu$_3$O$_{7-\delta}$ thin films at 4.2K.  In zero field, we observed conductance spectra with spontaneously-split peaks as well as spectra with unsplit peaks.  The former showed enhanced splitting with a $c$-axis field, and the latter showed threshold splitting above finite fields.  Although both field evolutions can be described in terms of screening and orbital supercurrents, within the framework of $d\pm i\alpha$ pairing, the enhanced splitting is consistent with only the $d$+$i\alpha$ state. The apparent absence of the $d$-$i\alpha$ state suggests that it is either energetically unfavorable in a field, or that the spontaneous peak splitting is due to an extrinsic mechanism other than intrinsically broken time-reversal symmetry. 

This work was supported by NSERC, CFI/OIT and the Canadian Institute for Advanced Research.


\begin{references}
\bibitem{Tunnelcone} By comparing the measured shift of the ASJ surface states with a quasiclassical model of the Doppler effect \cite{ZTH}, we estimate $\phi_c$ to be 15$^o-$30$^o$ in our STS junctions.
\end{references}
\end{document}